# On-Chip Multiplexed Multiple Entanglement Sources In a Single Silicon Nanowire

Yin-Hai Li,[3] Zhi-Yuan Zhou,[1,2,*] Lan-Tian Feng,[1,2] Wen-Tan Fang,[3] Shi-long Liu,[1,2] Shi-Kai Liu,[1,2] Kai Wang,[1,2] Xi-Feng Ren,[1,2] Dong-Sheng Ding,[1,2] Li-Xin Xu,[3] and Bao-Sen Shi[1,2]

[1]*CAS Key Laboratory of Quantum Information, USTC, Hefei, Anhui 230026, China*
[2]*Synergetic Innovation Center of Quantum Information & Quantum Physics,*
*University of Science and Technology of China, Hefei, Anhui 230026, China*
[3]*Department of Optics and Optical Engineering, University of Science and Technology of China,*
*Hefei, Anhui 230026, China*

*\*zyzhouphy@ustc.edu.cn*

The silicon-on-chip (SOI) photonic circuit is a very promising platform for scalable quantum information technology for its low loss, small footprint, and its compatibility with CMOS as well as telecom communications techniques. Multiple multiplexed entanglement sources, including energy-time, time-bin, and polarization-entangled sources based on 1-cm-length single-silicon nanowire, are all compatible with the (100-GHz) dense-wave-division-multiplexing (DWDM) system. Different methods, such as two-photon interference as well as Bell-inequality and quantum-state tomography, are used to characterize the quality of these entangled sources. Multiple entanglements are generated over more than five channel pairs with high raw (net) visibilities of around 97% (100%). The emission spectral brightness of these entangled sources reaches $4.2\times10^5$/(s.nm.mW). The quality of the photon pair generated in continuous and pulse pump regimes are compared. The high quality of these multiplexed-entanglement sources makes them very promising for use in minimized quantum communications and computation systems.

## I. INTRODUCTION

As a key resource in quantum information science, entanglement has broad applications in quantum communications, quantum computation, quantum-enhanced metrology, and in the study of fundamental physics in quantum mechanics [1, 2]. Usually entanglement can be generated from nonlinear processes such as spontaneous parametric down conversion (SPDC) [3–5] and spontaneous four wave mixing (SFWM) [6–10] in various physical systems. Because energy, momentum, and angular momentum are conserved, entanglement can be established between photons in their different degrees of freedoms. Some commonly used degrees of freedom used with photonic entangled states are polarization, energy-time, time-bin, and orbital angular momentum.

Although various systems can be used to prepare different entanglement states, the silicon-on-insulator (SOI) is a very promising platform, as it offers high integrating density and COMS compatibility. Based-on SFWM in SOI waveguides, much progress has been made to generate various non-classical photon-pair sources for various quantum information applications [8–17]. The noise level of SOI-based photon sources is much lower than dispersion-shifted fiber-based photon pair sources, because of a much narrower Raman scattering peak (105 GHz). This is in contrast to silica (10 THz) and therefore makes the SOI platform superior in building integrated quantum devices. In addition, SOI is also compatible with telecom communication techniques such as dense wave division multiplexing (DWDM), on which there have been some notable work [10, 15]. DWDM-based energy-time and time-bin entangled sources using the Si-micro cavity have been reported [9，10]. The internal pair-generation rate is very high because of cavity-enhanced effects. In contrast, a single-SOI nanowire waveguide has a broad photon-pair emission spectrum (several THz), which is well-suited for distribution over 100-GHz DWDM channels, thus offering great capacity for high-density quantum key distribution systems [18]. Also, the broad emission spectrum of photon pairs indicates a very short coherence time between photon pairs, which may have potential applications in quantum clock synchronization [19]. Although individual entangled photon sources or multiplexed sources based on SOI waveguides have been reported, no systematic study has been performed of multiplexed multiple entangled sources based on a single silicon nanowire.

In this article, multiple multiplexed entanglement sources, including energy-time, time-bin and polarization entangled sources based on a 1-cm-long single silicon nanowire, are shown to be compatible with (100-GHz) dense-wave-division-multiplexing (DWDM) systems. Different methods, such as two photon interference, and the Bell-inequality and quantum-state tomography, are used to characterize the quality of these entangled sources. Multiple entanglements are generated over more than 5 channel pairs with high raw (net) visibilities around of 97 % (100%). The emission spectral brightness of these entangled sources reaches $4.2\times10^5$ /(s.nm.mW). Also, the performance such as coincidence-to-background ratio (CAR), the pump-power dependence of CAR, and the dependence of CAR on the shift in wavelength from the pump wavelength of the photon pair generated in continuous and pulse pump regimes, are compared. The high quality of these multiplexed entanglement sources suggests possible use in minimized quantum communication and computation systems.

## II. MULTIPLEXED ENERGY-TIME ENTANGLED SOURCE

The silicon chip employed in our experiment is a single-silicon nanowire waveguide, which has a length of 1 cm and transverse dimensions of 220 nm (height)×450 nm (width). The external laser is coupled to the waveguide through gratings. The waveguide has an internal transmission loss of 2 dB/cm, the total insertion loss of the waveguide is 11 dB, including 4.5-dB coupling loss each at the input and output gratings.

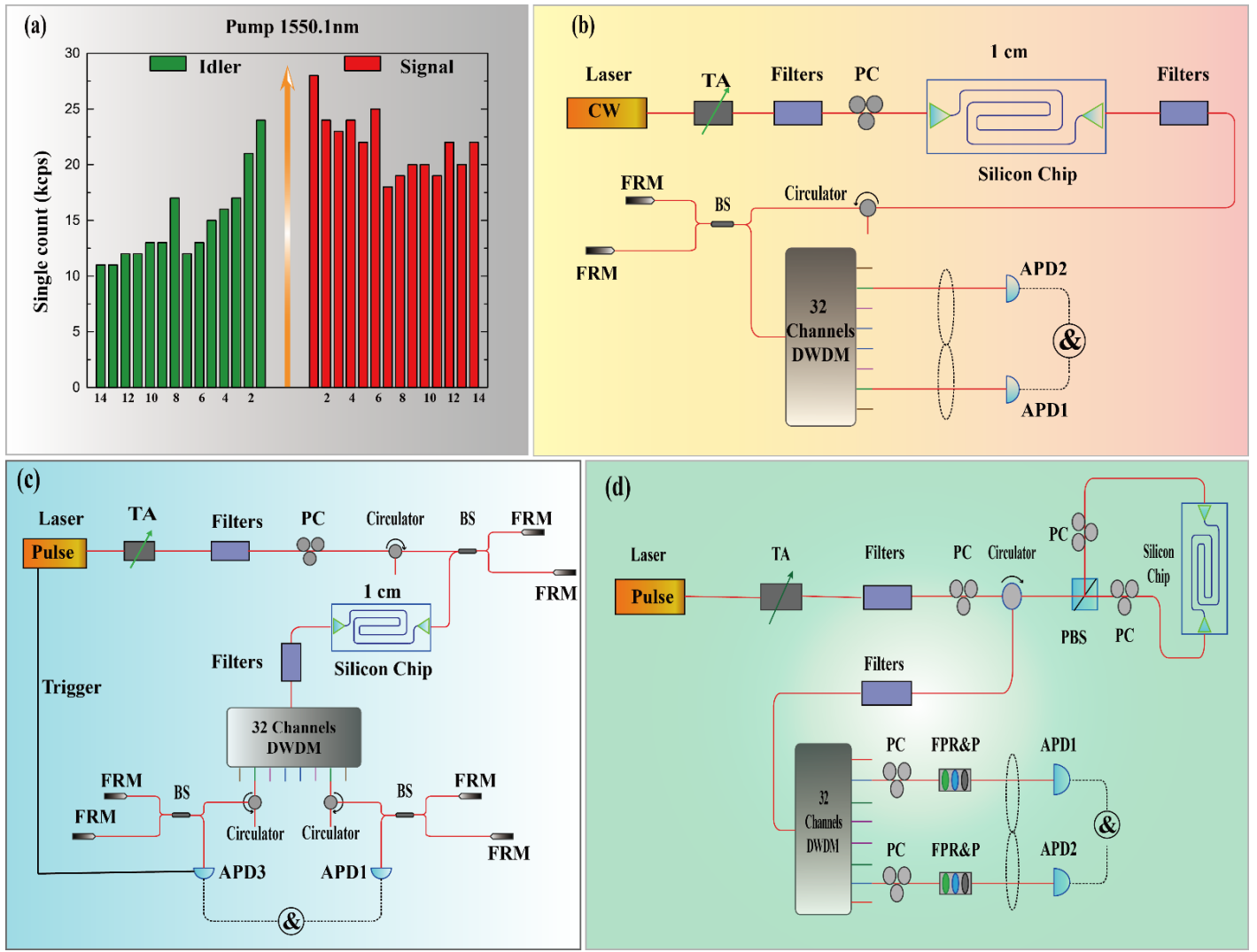

FIG. 1. Diagram showing the generation of multiplexed photon pair from SFWM, (a) the continuous emission spectral of the signal and idler photon is filtered by 32-channel DWDM system, the photon pairs which are equal depart from the central pump frequency are correlated. Experimental setups for multiplexed entanglement generation and analysis: (b) energy-time scheme (c) time-bin scheme, and (d) polarization. scheme. TA: tunable attenuator; BS: beam splitter; PC: fiber polarization controller; PBS: polarization beam splitter; FRM: Faraday rotation mirror; APD1, 2: free-running InGaAs avalanched single-photon detectors; APD3: gated-mode InGaAs avalanched single-photon detector; FPR&P: fiber polarization rotator and polarizer.

In the experimental setup [Fig. 1(b)], a tunable continuous wave (CW) narrow-bandwidth 1550.18-nm diode laser is first filtered by 100-GHz cascade DWDM filters to remove the broad-band background fluorescence light. The cleaned pump laser is injected into the silicon chip via an input coupling grating. Photon pairs are generated from the chip through the SFWM, and the photons coupling out from the chip are filtered by 200-GHz cascade filters with a rejection loss of more than 110-dB for the pump beam. The signal and idler photons pass through a common unbalanced Michelson interferometer (UMI, 1.6-ns time difference) for energy-time entanglement analysis [20]. The signal and idler photons from the output of the UMI are separated by a 32-channel DWDM device. The photons are detected using two free running InGaAs avalanche photon detectors (APD1, APD2, ID220, detection efficiencies 20%, dead time 5 μs, dark count, 3 kcps). By post-selection of the central peak of the coincidences with a proper time-window, the state after post-selection is

$$|\Phi\rangle_1 = \frac{1}{\sqrt{2}}(|SS\rangle + |LL\rangle), \quad (1)$$

where the first and second terms denote both signal and idler photons passing the short ($S$) and long ($L$) arm of the UMIs.

The photon pair generated from the silicon waveguide has a continuous broadband emission spectrum [Fig. 1(a)] for the signal and idler photons. By filtering the signal using a 32-channel DWDM, the corresponding channels at equal shifts from the pump frequency are correlated [21]. For simplicity, the standard ITU grids are relabeled $i1$–$i14$ and $s1$–$s14$. For a detailed definition of the central wavelength of the ITU grid, see Table A1 in Appendix A. Fig. 1(a) shows the single count rates of the signal and idler photons for different DWDM channels. The differences in single count rates arise from the different SFWM gains, Raman scattering of the signal and idler bands, and non-uniform losses of different DWDM channels. We first characterize the photon pair source by measuring CARs over different correlated channel pairs. The experimental results are shown in Fig. 2(a). The 14 channel pairs are measured from $i1$–$s1$ to $i14$–$s14$. The CAR increases when the shift in the signal and idler wavelength increases. The coincidence time-window is 0.8 ns, and the cleaned pump power before injection into the chip is 1.37 mW. The average coincidence is about 51 per second, therefore the emission spectral brightness reaches $4.2 \times 10^5$/(s.nm.mW) after output coupling losses (5.5 dB), filtering losses (4 dB), and detection losses of the detectors for each photon are taken into account.

Next we characterized the photon source by measuring CAR as a function of the pump power for $i8$–$s8$ [Fig. 2(b)]. The CAR increases in the low pump-power regime and decreases in the high pump-power regime. At low pump power, the CAR is dominated by dark coincidences from the detectors and noise photons from the filters. At high pump power, the CAR is dominated by multi-photon emission events. The fitted curve is obtained using the method introduced in [22, 23].

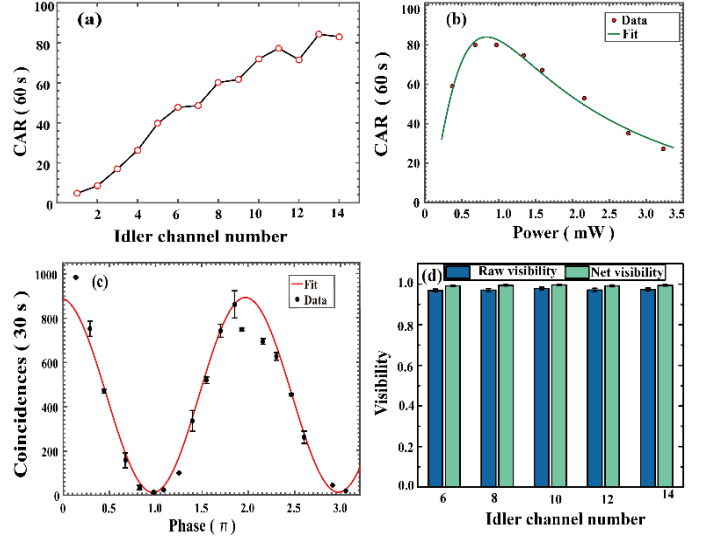

FIG. 2. (a) CARs for different correlated channel pairs from $i1$ to $i14$. (b) CARs as a function of pump power. (c) Coincidences in 30 s as a function of the total phase for the signal and idler channels $i8$ and $s8$. (d) Raw and net visibilities for 5 correlated channel pairs ranging from $i6$ to $i14$.

To demonstrate the high quality of the energy-time entanglement source, we measured the two-photon interference fringes between the two photons for channel pairs $i8$–$s8$. The result [Fig. 2(c)] is given as coincidence counts in a 30-s interval as a function of the measured UMI phase ($\phi_s + \phi_i$). The interference fringe has a raw (net) visibility of 97.10 ±0.88% (99.08 ±0.48%). The UMI used in the experiments is the same as in [7]; it is individually sealed in a copper box and thermally insulated from the air. The temperature of each copper box is controlled using homemade semiconductor Peltier temperature controller with temperature fluctuations of ±2 mK. For details of the UMI characteristics, see Appendix C. To quantify the multi-channel performance of the energy-time entangled source, we measured the raw visibilities for 5 correlated channel pairs from $i6$–$s6$ to $i14$–$s14$ [Fig. 2(d)]; the raw (net) visibilities are all around 97% (100%). The high visibilities obtained for multi-channel pairs indicate high quality of the source of greater than 71% visibility, which also implies the presence of Bell non-locality between the signal and idler photon [24].

## III. MULTIPLEXED TIME-BIN ENTANGLED SOURCE

Next, we characterize the multiplexed time-bin entangled source. The time-bin entangled photon-pair source is another important quantum resource suitable for performing quantum key distribution over telecom fibers. As in previous experiments, the scheme for the time-bin entanglement generation [Fig. 1(c)] involves using the mode-locked fiber pulse laser (25 ps) and forward and backward filters. The pump beam is split into two time bins using a 1.6-ns UMI consisting of a fiber coupler and two Faraday rotator mirrors. Each pump time-bin generates two photon pairs. After removing the pump beam, the generated photon pairs are separated by a 32-channel 100-GHz DWDM. The correlated channel pairs are connected to two UMIs with the same parameters as the first UMI. As for the UMI above, each UMI is sealed, thermally insulated, and temperature controlled to within fluctuations of ±2 mK. The output of one of the interferometers is connected to APD3 (Princeton Lightwave Inc., Cranbury, NJ, USA, 100-MHz trigger rate, 1-ns detection window, 15% detection efficiency), which is gated using the synchronize signal from the mode-locked laser. The output from the other interferometer is connected to APD1. The electrical detection output signals from APD1 and APD3 are sent for coincidence measurements.

We give a brief theoretical description of our time-bin entanglement source. After the pump beam is divided into two time slots by forward UMI, the pump photon is prepared in state $|\Psi\rangle_p = 1/\sqrt{2}(|S\rangle - e^{i\phi_p}|L\rangle)$, where $\phi_p$ is the relative phase between the two arms. After transmitting the pump beam from the silicon waveguide, time-bin entanglement is created from the two time slots with the states expressed as [25]

$$|\Phi\rangle = \frac{1}{\sqrt{2}}(|SS\rangle - e^{i2\phi_p}|LL\rangle). \quad (2)$$

When this entanglement state is further transformed by two UMIs in the signal and idler ports, and we post-select the central slot, two-photon interference fringes are obtained. The coincidence for the two photons is proportional to $1 - V\cos(2\phi_p - \phi_s - \phi_i)$, where $\phi_s$ and $\phi_i$ are the relative phases of the UMIs in the signal and idler ports, respectively.

We measure first the CARs for different channel pairs and the CARs as a function of the pump power in the pulse pump regime. The results are shown in Fig. 3. From Fig. 3(a), the pump power is 53 µW and the coincidence window is 0.4 ns. By comparing with the CW pump regime shown in Fig. 2(a) and (b), we find the same basic behaviors for both CW and pulse pump regimes. The differences are as follows: because the effect of Raman scattering is reduced in the pulsed regime, the noise is lower in the pulse regime, which leads to CARs that are larger than for the CW pump regime. Because the peak power is much higher in the pulsed regime, the generation rate for multi-photons will be much higher than for the CW pump regime given same average power. This is the reason for a greater decreasing rate for CARs in the pulse pump regime [Fig. 3(b)] than in the CW pump regime [Fig. 2(b)].

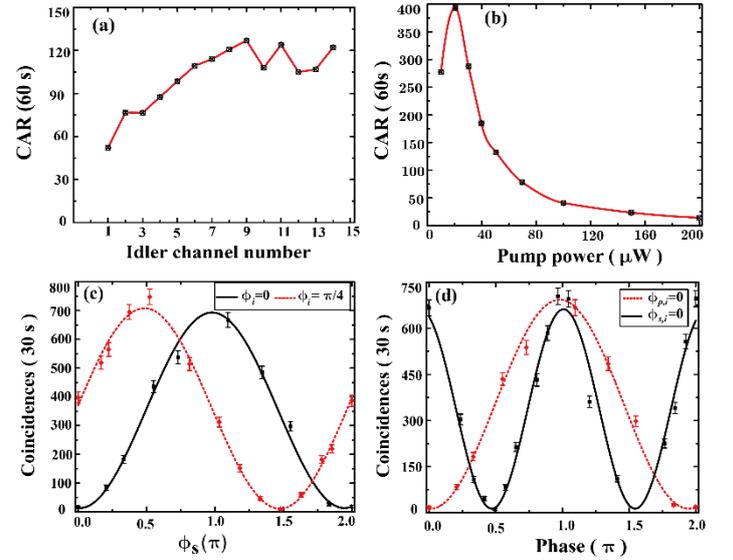

FIG. 3. CARs as a function of (a) correlated channel pairs $i1$-$i14$ and (b) average pump power for channel $i8$. Coincidences over 30-s intervals as a function of: (c) the signal UMI phase when the idler UMI phase is set at 0 and $\pi/4$, and (d) the signal UMI phase (dashed red line) and the pump UMI phase (solid black line).

To characterize the quality of our time-bin entangled source in detail, we first measured the dependence of the two-photon interference with respected to the pump, signal, and idler UMIs phases ($\phi_p$, $\phi_s$, $\phi_i$) by tuning the temperatures of the three UMIs. From the results shown in Fig. 3(c), the pump phase is set to 0, while the idler phase is set to 0 (solid black curve) and $\pi/4$ (dashed red curve). The raw (net) visibilities are 96.31±0.99% (99.36±0.69%) and 97.09±1.70% (98.92±1.55%). The two curves in panel (d) show the results for $\phi_p = \phi_i = 0$ (dashed red line) and $\phi_s = \phi_i = 0$ (solid black line) verifying that the two-photon interference fringe has a period of oscillation of $\pi$ for the pump phase and $2\pi$ for the signal (idler) phase.

For the multi-channel performance of the time-bin entangled source (Fig. 4), we measured visibilities for five correlated channel pairs from $i6$–$s6$ to $i14$–$s14$. The raw (net) visibilities are up around 96% (100%). The high visibilities obtained for multi-channel pairs indicate a high-quality source, which also implies the presence of Bell non-locality between signal and idler photons.

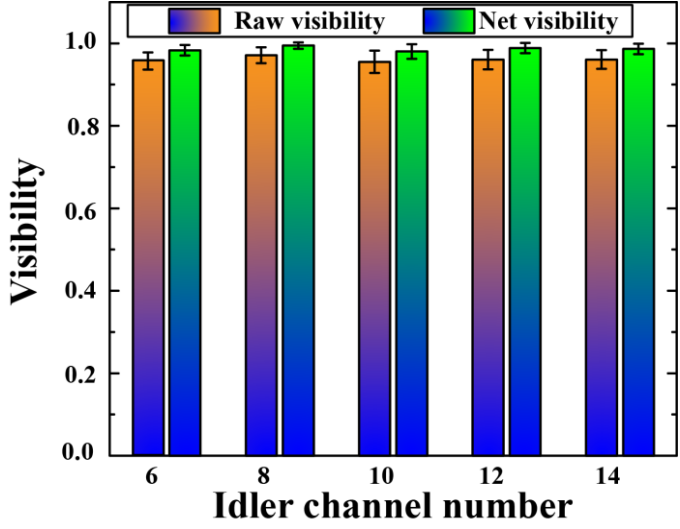

FIG. 4. Raw and net visibilities for signal and idler pair of $i6$ to $i14$.

## IV. MULTIPLEXED POLARIZATION ENTANGLED SOURCE

Finally, we shall characterize the multiplexed polarization entangled source. The experimental setup for generating polarized entangled photon source is shown in Fig. 1(d). The pre-filter and post filter for the pump beam and the photon pair are the same as used in the energy-time entanglement generation. After removing the strong pump beam, a 32-channel 100-GHz DWDM is used for distributing the polarization entanglement of correlated channel pairs. The output ports of each channel pair are connected to APDs. The output signals from ADP1 and ADP2 are sent to our coincidence count device. The principle for polarization entanglement generation is similar to a Sagnac interferometer loop for generating polarization entangled photons based on second-order nonlinear crystals in the SPDC processes [4].

In the present instance, the pump beam is split by a polarized beam splitter (PBS) into clockwise (CW) and counterclockwise (CCW) circulation directions. As the waveguide only sustains one mode, by adjusting the polarization controllers PC1 and PC2 in the loop interferometer, the power coupled to the waveguide can be maximized. Therefore, in the CW direction, the photon pair generated are $|H\rangle_s|H\rangle_i$. After rotation by PC2, the polarization changes to $|V\rangle_s|V\rangle_i$. While in the CCW direction, the photon pair with polarization state $|H\rangle_s|H\rangle_i$ remains unchanged at the PBS. After the photon pairs from the two counter propagation directions recombined at the PBS, the photonic state at the output port of the Sagnac loop can be expressed as [26]

$$|\Phi\rangle = |HH\rangle + \eta e^{i\delta}|VV\rangle, \quad (3)$$

where $\eta$ is determined by the ratio of the pump power and $\delta = 2(\varphi_p + \Delta k_p L)$ depends on the initial phase $\varphi_p$ of the pump beam at the input port and the birefringence $\Delta k_p = k_{pH} - k_{pV}$ experienced by the pump beam in the $H$ and $V$ polarizations. By changing the pump power and initial phase, we can generate maximally entangled states $|\Phi\rangle^\pm = 1/\sqrt{2}(|HH\rangle \pm |VV\rangle)$.

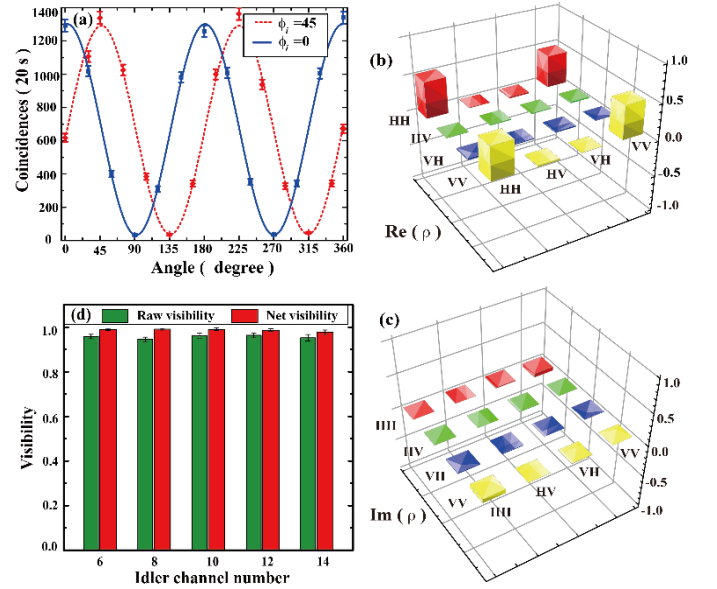

FIG. 5. (a) Coincidences in 20-s intervals as a function of the signal polarizer angle, when the idler polarizer is keeping at 0 degree (solid blue line) and 45 degree (dashed red line), respectively. (b), (c) Real and imaginary parts of the reconstructed density matrix. (d) Raw and net visibilities for signal and idler pairs ranging from $i6$ to $i14$.

To characterize the quality of our entangled source, different methods are used in our experiment. We first measured the two-photon polarization interference fringes under different settings. From the experimental results [Fig. 5(a)], the raw (net) visibilities in the 0 degree (solid blue line) and 45 degree (dashed red line) bases for signal and idler pair $i8$ and $s8$ are 95.16±1.23% (99.11±1.12%) and 95.41±1.79% (99.26±1.27%), respectively. Visibilities above 71% imply the presence of quantum entanglement.

To further characterize the performance of the entangled source, we measured the $S$ parameter of the CHSH

inequality [27], which for the two sets of polarization direction settings ($\theta_s$ =−22.5, 67.5, 22.5, 112.5 degree; $\theta_i$ =−45, 45, 0, 90 degree) is 2.66±0.10 without background subtraction, violating the inequality at more than 6 standard deviations.

To know precisely which Bell state is generated, we also performed quantum tomography. The experimental density matrix $\rho_{exp}$, reconstructed using the maximum-likelihood estimation method [28], is displayed in Fig.5(b) and (c). The fidelity of the reconstructed density matrix to the ideal Bell state $|\Phi\rangle$ is defined as $F = [Tr(\sqrt{\sqrt{\rho_{th}}\rho_{exp}\sqrt{\rho_{th}}})]^2$, where $Tr$ is the trace, $\rho_{th}$ is the ideal density matrix. We estimated the fidelity of our present source at 0.934±0.015. Therefore the generated polarization entangled state is nearly a maximal entangled Bell state. The deviation of the fidelity from unity is mainly due to dark coincidences and errors in rotating the angle of the wave plates.

The multi-channel performance of the polarization entangled source [Fig. 5(d)] shows high raw and net visibilities in the 45° basis for all measured signal and idler pairs.

## IV. CONCLUSION AND OUTLOOK

We have realized high-quality multiplexed multiple entanglement sources based on a single SOI nanowire waveguide. These sources are characterized by various methods. Both CW pump and pulse pump regimes were studied and compared. The performance of the sources can be further improved using better single-photon detectors such as super-conducting nanowire single-photon detectors with higher efficiency, lower dark counts and lower time jitter [29]. These sources are fully compatible with the C-band DWDM technique, which may be of great importance for future high-capacity quantum communications systems.

By comparing the present entangled sources based on single silicon nanowire with micro-cavity-based photon entangled sources [9, 10, 15], the quality of entanglement measured from the visibility of interference fringes are nearly the same. A micro-cavity-based source has higher brightness because of cavity-enhanced effects than single silicon nanowire based sources given the same pump power. The bandwidth of the micro-cavity based sources is much narrower than the present source, which requires fine tuning the pump wavelength to the resonance peak of the cavity. The temperature of the chip also needs active stabilization to avoid wavelength drifts of the resonance peaks.

In table 1, we give a detail comparison of the present source in CW pump regime with other waveguide sources based on second- or third-order nonlinear processes, the data in table 1 is taken from ref. [16]. It is obvious that second order nonlinear PPLN waveguide has a brighter spectral brightness than third order nonlinear waveguide [30], but it is not CMOS compatible and the CAR is relative lower than Si waveguide. The CMOS compatible second order nonlinear AlGaAs waveguide [31] has nearly the same order of brightness as third order nonlinear Si waveguide [12], but has relative lower CAR. The cavity-based Si waveguide sources have much brighter spectral brightness than single pass Si waveguide because of cavity enhanced effects for both the pump beam and the emitted photon pair [16, 32], which lead to the same order of brightness as PPLN waveguide. The spectral brightness of our source is at the same order of magnitude as refs. [12, 31], but the CAR are higher than these two works.

Table 1. Comparison between integrated waveguide sources based on different materials and nonlinear processes at room temperature, the data are taken from ref. [16].

| Reference | Structure | Material | CAR | Spectral brightness (P=1mW, $s^{-1}$ $nm^{-1}$) |
|---|---|---|---|---|
| [30] | Waveguide | PPLN | ~6 | ~7.5×$10^7$ |
| [31] | Waveguide | AlGaAs | ~7 | ~6×$10^5$ |
| [12] | Waveguide | Si | ~30 | ~4×$10^5$ |
| [32] | CROW | Si | ~8 | ~3×$10^6$ |
| [16] | μ ring | Si | ~64 | ~6×$10^7$ |
| This work | Waveguide | Si | ~80 | ~4.2×$10^5$ |

By applying quantum frequency conversion [33], these sources will be very useful to link with other quantum systems such as quantum memories. By integrating more functionality on the chip such as on-chip pre-filters for the pump laser background fluorescence and post-filters to reject the pump beam, the chip could act as a minimized entanglement photon emission device [17].

## ACKNOWLEDGMENTS
This work is supported by National Natural Science Foundation of China (Grant Nos. 11174271, 11604322, 61275115, 61435011, 61525504, and 61605194), China Postdoctoral Science Foundation (Grant No. 2016M590570) and the Fundamental Research Funds for the Central Universities.

## REFERENCES
[1] J.-W. Pan, Z.-B. Chen, C.-Y. Lu, H. Weinfurter, A. Zeilinger, and M. Zukowski, Multiphoton entanglement and

## Appendix

### A. Definition of the wavelengths of the standard ITU grids

The corresponding wavelengths for correlated signal and idler photons are defined in Table A1, the bolded channels are used for entanglement measurements of the multi-channel performance in the experiments. The pump wavelength is located at the central of channel C34.

Table A1. Definition of the wavelengths of the standard ITU grids for the signal and idler photons.

| Pair number | DWDM channel | Wavelength ( nm ) |
|---|---|---|
| **Signal 14 - Idler 14** | **C19 - C49** | **1562.23 - 1538.19** |
| Signal 13 - Idler 13 | C20 - C48 | 1561.42 – 1538.98 |
| **Signal 12- Idler 12** | **C21 – C47** | **1560.61-1539.77** |
| Signal 11 - Idler 11 | C22 – C46 | 1559.79 – 1540.56 |
| **Signal 10 - Idler 10** | **C23 – C45** | **1558.98 – 1541.35** |
| Signal 9 - Idler 9 | C24 – C44 | 1558.17 – 1542.14 |
| **Signal 8 - Idler 8** | **C25 – C43** | **1557.36 – 1542.94** |
| Signal 7 - Idler 7 | C26 – C42 | 1556.56 – 1543.73 |
| **Signal 6 – Idler 6** | **C27 – C41** | **1555.75 – 1544.53** |
| Signal 5 – Idler 5 | C28 – C40 | 1554.94 – 1545.32 |
| Signal 4 - Idler 4 | C29 – C39 | 1554.13 – 1546.12 |
| Signal 3 - Idler 3 | C30 – C38 | 1553.33 – 1546.92 |
| Signal 2 - Idler 2 | C31 – C37 | 1552.52 – 1547.72 |
| Signal 1 - Idler 1 | C32 –C36 | 1551.72 – 1548.52 |
| **Pump** | **C34** | **1550.12** |

### B. More data on energy-time entangled source

Generally, we can increase the CAR in coincidence measurement by reducing the time-window for the measurement. This effect is shown in Fig. A1(a), we can see that the CAR is decreasing very fast when the coincidence window is increasing. This phenomenon indicates that the CAR can be further increasing if low time jitter single photon detector is used in experiments. Fig. A1(a) shows the CAR measured for signal and idler channel $s8$ and $i8$, the CAR reaches about 150 for coincidence window of 0.4 ns, and decreases to about 20 for coincidence window of 3.2 ns.

Fig. A1(b) shows single count rates as a function of the pump power for channel $s8$ and $i8$, single count rates are increasing squared with the increasing of the pump power. Differences in count rates between channel $s8$ and $i8$ are because of Raman scattering photons come from pigtails of filters, this difference can be reduced by using pulse pump laser. The data are fitted based on method introduced in ref. [34]

Fig. A1(c), and (d) are two specific coincidence histograms in 30 s for energy-time entanglement measurement. The coincidence window is 0.4 ns in the measurement. The corresponding total phase $\phi_s + \phi_i$ for the two cases are π/2 and 0, respectively. We can see that destructive interference in Fig. A1(c) is at the level of background coincidences.

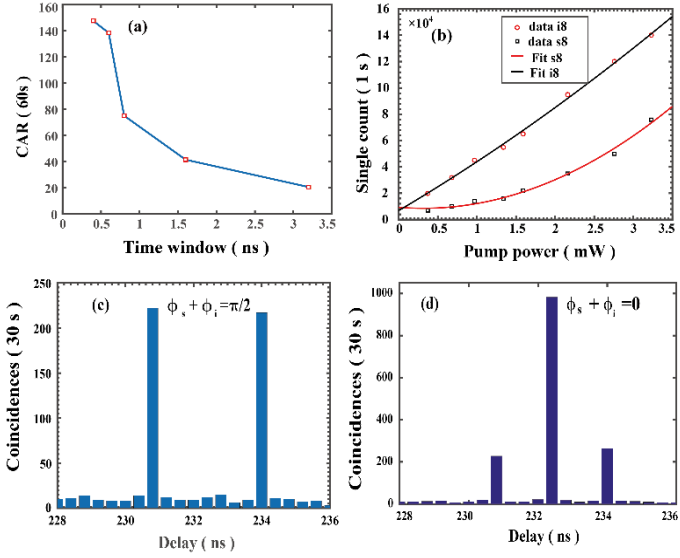

Fig. A1. (a) CAR as a function of coincidence window. (b) single count rates as a function of pump power for signal and idler channels $s$8 and $i$8. (c), (d) coincidences histograms in 30 s for total signal and idler phase of π/2 and 0, respectively.

### C. How the UMI phase is tuned in time-bin source

The thermal coefficient of fiber at 1550 nm is $\frac{dn}{dT} = 0.811 \times 10^{-5} /^0 C$, the fiber length difference of the UMI is (163.48 mm) $L_d = c\Delta t / 2n$ for 1.6 ns delay. The temperature for one tuning period $\Delta T = \lambda / (2L_d \frac{dn}{dT})$ is 0.585K. In the experiments, the temperature tuning periods and the phase of the UMIs are measured and calibrated using a stable narrow bandwidth laser source. The phase of the UMIs can keep unchanging for hours because of seriously thermal and acoustic isolation from the environment.

### D. The power tuning behaviors of the polarization entangled source

Fig. A2 shows the power detuning behaviors of the polarization entangled source. Fig. A2(a) shows the signal count rate of channel $i$8 as a function of average pump power in the pulse pump regime. Fig. A2(b) shows the power dependence of the visibility in 45 degree basis, it shows that the raw and net visibility is decreasing when the pump power is increasing. The decreasing of the visibility is mainly caused by increasing of multi-photon emission events in the high pump power regime.

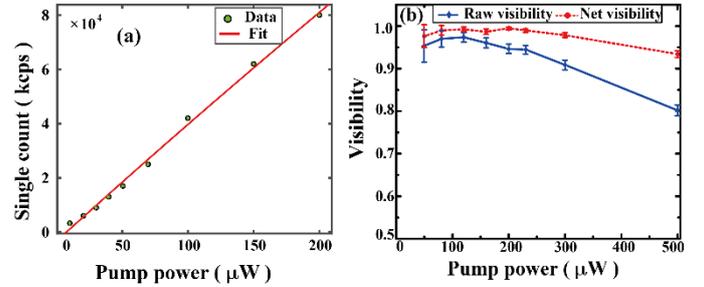

Fig. A2. (a). signal count rate of channel $i$8 as a function of average pump power in the pulse pump regime (b) power dependence of the visibility in 45 degree basis.